\documentclass[]{iopart}
\usepackage{iopams}  
\usepackage{psfig}  
\usepackage{times}

\begin{document}

\letter{Rossby-Haurwitz waves of a slowly and
	differentially rotating fluid shell}

\author{
Luciano Rezzolla$^{1,2}$ and Shin'ichirou Yoshida$^1$}

\address{$^1$SISSA, International School for Advanced Studies,
        Via Beirut 2, 34014 Trieste, Italy}

\address{$^2$INFN, Department of Physics, University of
        Trieste, Via Valerio, 2 34127 Trieste, Italy}

\begin{abstract}
Recent studies have raised doubts about the occurrence of
$r$ modes in Newtonian stars with a large degree of
differential rotation. To assess the validity of this
conjecture we have solved the eigenvalue problem for
Rossby-Haurwitz waves (the analogues of $r$ waves on a
thin-shell) in the presence of differential rotation. The
results obtained indicate that the eigenvalue problem is
never singular and that, at least for the case of a
thin-shell, the analogues of $r$ modes can be found for
arbitrarily large degrees of differential rotation. This
work clarifies the puzzling results obtained in
calculations of differentially rotating axi-symmetric
Newtonian stars.
\end{abstract}


\section{Introduction}

	The investigation of $r$ modes in rotating stars
has seen a renewed interest in recent years since they
were shown to be unstable to the emission of
gravitational waves (Andersson 1998, Friedman and Morsink
1998). In the last three years a number of papers have
outlined the basic properties of the $r$-mode
oscillations, their role in triggering and feeding the
instability, and the importance of this instability in
the emission of gravitational waves as well as in the
slowing down of rapidly rotating, hot neutron stars (see
Andersson and Kokkotas 2001 and Friedman and Lockitch
2001 for some recent reviews).  Recent studies, on the
other hand, have focussed on more intricate but also more
realistic aspects of the instability.  Such aspects
include the interaction of the instability with magnetic
fields (Spruit 1999, Rezzolla et al. 2000), nonlinear and
secular effects (Rezzolla et al. 2001a, 2001b;
Stergioulas and Font 2001; Lindblom et al.  2001), the
presence of differential rotation in Newtonian stars
(Karino et al. 2001, hereafter KYE), of a continuous
frequency spectrum in the slow-rotation approximation of
uniformly rotating relativistic stars (Kojima and
Hosonuma 1999) and the lack of discrete, physically
plausible mode solutions within the continuous part of
the spectrum (Ruoff and Kokkotas 2001).  We here pay
attention to one of these more ``subtle'' aspects and
focus on the role played by differential rotation in
Newtonian stars. In particular, we address the question
of whether a large degree of differential rotation in
slowly rotating stars turns the eigenvalue problem for
$r$ modes into a singular eigenvalue problem, thus
preventing the existence of $r$ modes.  This is an
important question because of the possible observation of
a singular problem in $r$-mode oscillations that has
emerged from recent calculations of differentially
rotating axi-symmetric Newtonian stars (KYE). More
intriguingly, similar evidence is converging also from
different approaches such as those looking at purely
axial $r$ modes in rotating non-isentropic relativistic
stars in the slow-rotation approximation (Yoshida 2001,
Yoshida and Futamase 2001; Ruoff and Kokkotas, 2001a,
2001b) although a regularization may be possible in some
cases (Lockitch and Andersson 2001).

	Part of the subtlety introduced by differential
rotation in the eigenvalue problem is due to the fact
that, above a certain degree of differential rotation,
the eigenvalue equations become very stiff as a result of
large radial and polar gradients of the rotational
angular frequency. Under these circumstances, the
numerical solution requires increasingly high accuracy
and computational costs. To circumvent this problem, we
have here resorted to a simpler model based on a
differentially rotating, thin shell. A thin-shell model
has been adopted also by Levin and Ushomirsky (2001) to
investigate the effect of electromagnetic
radiation-reaction for $r$ modes on a uniformly rotating
shell. This approach replaces the set of partial
differential equations of a multidimensional star model
with an ordinary differential equation which can be
solved to much higher accuracy and with modest
computational costs. Although much simpler to solve, the
$r$-mode eigenvalue problem for a thin shell incorporates
many of the mathematical properties of the corresponding
eigenvalue problems for multidimensional Newtonian stars
or for slowly-rotating relativistic stars. As will become
clear in the following Sections, adopting a thin-shell
model has been very valuable for gaining insight into the
behaviour of $r$ modes in differentially rotating fluids.

\section{Basic Equations}
\label{be}

	The thin-shell model adopted here is based on the
following simplifying assumptions: {\it (i)} the
background star is slowly rotating and we will omit terms
in the hydrodynamical equations of ${\cal O}(\Omega^2)$;
{\it (ii)} the fluid is incompressible and inviscid; {\it
(iii)} the shell is spherical with radius $R$ for every
rate of differential rotation; {\it (iv)} the fluid
motion is constrained on the shell, that is, we have no
radial component of velocity.

	In the inertial (nonrotating) frame the fluid
velocity $\vec{v}$ is the composition of the unperturbed
velocity of the differentially rotating shell
$\vec{v}_{_0}$, where, in spherical polar coordinates,
$v^i_{_0}=(0,0,\Omega(\theta) R \sin\theta)$, with the
velocity perturbation $\delta \vec {u}$ which has only
tangential components, i.e. $\delta u^i =
(0,u^{\theta},u^{\phi})$. As a result, the components of
the fluid velocity can be written as
\begin{equation}
v^i = (0, u^{\theta}, \Omega R\sin\theta + u^{\phi}),
\end{equation}

	In the simple model considered here, the fluid
motion in the shell is fully described by the Euler and
continuity equations. The latter, in particular, is
trivially satisfied by the background motion and assumes
the following form in its perturbed part
\begin{equation}
\partial_{\theta}(u^{\theta}\sin\theta) 
	+ \partial_{\phi}  u^{\phi} = 0 \ .
\label{continuity}
\end{equation}

	As customary with this type of problem, it is
here convenient to introduce the fluid vorticity
$\vec{\xi} \equiv\frac{1}{2}\nabla\times\vec{v}$, whose
radial component has the form
\begin{equation}
\xi^r = \Omega\cos\theta + \frac{1}{2}\frac{d\Omega}{d\theta}
	\sin\theta +\frac{1}{2r\sin\theta}
	\left[\partial_{\theta}(u^{\phi}\sin\theta) 
	- \partial_{\phi}  u^{\theta}
	\right] \ .
\end{equation}

	It is a simple exercise to show that the Euler
equation can then be cast into a conservation equation
for the vorticity $\partial_t\vec{\xi} + \nabla\times
(\vec{\xi}\times\vec{v}) = 0$, whose radial component is
\begin{equation}
\left( \partial_t + \frac{v^{\theta}}{r}
	\partial_{\theta} + \frac{v^{\phi}}{r\sin\theta}
	\partial_{\phi}\right)\xi^r = 0 \ .
\label{xi-r}
\end{equation}

	Because we are interested in the harmonic modes, 
we assume that the tangential components
of the perturbation velocity can be written as
$u^{\theta}, u^{\phi} \sim {\rm exp}(-i\sigma t +
im\phi)$, so that the linearized version of
equation~(\ref{xi-r}) is
\begin{equation}
\fl -i(\sigma-m\Omega)
	\left[\frac{d}{d\theta}(u^{\phi}\sin\theta)
	-imu^{\theta}\right]
	+ 2u^{\theta}\sin\theta \frac{d}{d\theta}
	\left(\Omega\cos\theta + \frac{1}{2}
	\frac{d\Omega}{d\theta}\sin\theta\right) = 0\ .
\end{equation}

	We can now use the continuity
equation~(\ref{continuity}) to rewrite the radial
component of the vorticity conservation equation as
\begin{equation}
\label{eigen_eq}
\fl \frac{d}{d\mu}\left[(1-\mu^2)\frac{d\chi}{d\mu}\right]
	- \frac{m^2}{1-\mu^2}\chi
	- \frac{2m\Omega}{\sigma-m\Omega}
	\left[
	1 + \left(\frac{2\mu}{\Omega}\right)\frac{d\Omega}{d\mu}
	- \left(\frac{1-\mu^2}{2\Omega}\right)
	\frac{d^2\Omega}{d\mu^2}\right]\chi = 0\ ,
\end{equation}
where we have introduced the new coordinate $\mu \equiv
\cos\theta$ and where $\chi \equiv u^{\theta}
\sin\theta$. Together with regular boundary conditions at
the northern and southern poles ($\mu=\pm 1$),
equation~(\ref{eigen_eq}) accounts for the eigenvalue
problem of the fluid normal modes on the shell. In the
case of a uniformly rotating shell,
${d\Omega}/{d\mu}=0={d^2\Omega}/{d\mu^2}$, and
equation~(\ref{eigen_eq}), first derived by Haurwitz in
1940 (see also Stewartson and Rickard, 1969), reduces to
Legendre's equation whose regular solution is
$\chi=P_{^l}^{_m}(\mu)$, with $P_{^l}^{_m}(\mu)$ being
the associated Legendre functions. Note that in the case
of uniform rotation the eigenfrequencies obey the well
known dispersion relation for $r$ modes
\begin{equation}
\label{dispersion}
\sigma = m\Omega - \frac{2m\Omega}{l(l+1)}
	\ ,
\end{equation}
and this justifies calling these modes Rossby-Haurwitz
waves.  For simplicity, however, hereafter we will refer
to them as $r$ modes.

	The solution of equation~(\ref{eigen_eq}) in the
case of a differentially rotating shell depends on the
law of differential rotation chosen and on its first and
second polar derivatives. The choice of a law of
differential rotation is, in this sense, somewhat
arbitrary but we have here followed previous work on the
subject (Eriguchi and M\"uller 1985, KYE) and modeled the
differential rotation through either a ``$j={\rm
const.}$'' law
\begin{equation}
\label{diff_rot_j}
\Omega = \frac{A^2_j + 1}{A^2_j + 1 - \mu^2}
	\Omega_{_{\rm E}} \ ,
\end{equation}
or a ``$v={\rm const.}$'' law\footnote{These differential
laws take their names from the fact that through the
variation of the parameter $A_{j,v}$ they represent
families of rotation laws spanning the range between
uniform rotation (for $A_{j,v}\to\infty$) and
differential rotation with constant specific angular
momentum or constant linear velocity (for $A_{j,v}\to
0$), respectively.}
\begin{equation}
\label{diff_rot_v}
\Omega = \frac{A_v + 1}{A_v + \sqrt{1 - \mu^2}}
	\Omega_{_{\rm E}}\ .
\end{equation}
In both cases $\Omega_{_{\rm E}}$ is the angular velocity
at the equator ($\mu=0$) and the parameter $A_{j,v} > 0$
accounts for the degree of differential rotation so that
uniform rotation (i.e. ${\Omega}/{\Omega_{_{\rm E}}}=
{\rm const.}$) is reached for $A_{j,v} \to \infty$. Note
that the use of a law of differential rotation is
physically plausible as long as such a law does not
violate Rayleigh's stability criterion for rotating
inviscid fluids: $d\left(\varpi^2
\Omega\right)^2/{d\varpi} > 0$, where $\varpi$ is the
cylindrical radial coordinate. It is straightforward to
check that both expressions (\ref{diff_rot_j}) and
(\ref{diff_rot_v}) satisfy Rayleigh's criterion for all
values of $A_{j,v} > 0$.

	The solution of equation~(\ref{eigen_eq}) is
considerably simpler than the solution of the
corresponding set of partial differential equations
describing normal modes of a differentially rotating
axi-symmetric star and discussed by KYE. Nevertheless,
most of the mathematical properties that are found in the
solution of the system of partial differential equations
can already be found in (\ref{eigen_eq}). A particularly
important property of equation~(\ref{eigen_eq}), which
has been encountered also when dealing with $r$ modes of
differentially and rapidly rotating stars (KYE), is that
it may become a singular eigenvalue problem at the
angular position $\mu_s$ for which
\begin{equation}
\label{corot}
{\sigma-m\Omega(\mu_s)}= 0 = \omega_{\rm ph} -
	\Omega(\mu_s) \ ,   
\end{equation}
where $\omega_{\rm ph} \equiv \sigma/m$ is the phase
velocity of the mode. The condition (\ref{corot}), which
can be interpreted as the appearance of a {\it
``corotation point''} (i.e. a point on the shell at which
the perturbation pattern rotates at the same angular
velocity as the background shell) has been interpreted by
KYE as the cause impeding the numerical solution of the
eigenvalue problem for a sufficiently large degree of
differential rotation\footnote{KYE also noted that the
difficulty in finding a convergent numerical solution was
dependent on the rate of rotation of the background
models, with rapidly rotating models providing solutions
for comparably smaller values of $A_{j,v}$.}.
Interestingly, equation~(\ref{eigen_eq}) offers close
analogies also with the corresponding equation obtained
in full General Relativity for a slowly and uniformly
rotating relativistic star (Yoshida 2001, Yoshida and
Futamase 2001; Ruoff and Kokkotas, 2001a, 2001b). The
analogy is brought about by the fact that the corrections
due to the relativistic dragging of inertial frames are
mathematically similar to the corrections due to
differential rotation and introduce a similar coefficient
which could become singular for certain rates of
rotation\footnote{Note that although
equation~(\ref{eigen_eq}) can in principle be singular, a
global solution can still be found if series expansion
techniques, such as the Frobenius method, are employed in
the vicinity of the singular point $\mu_s$.}.

\section{Strategy of the Numerical Solution}

	The numerical procedure adopted in the solution
of equation~(\ref{eigen_eq}) is straightforward and needs
particular care only if a singular point should appear
during the solution. In general, we can numerically
integrate equation~(\ref{eigen_eq}) as a two point
boundary value problem with boundary conditions at both
poles, $\mu=\pm 1$, being given by the requirement that
the solution $\chi$ is regular there. The two solutions
found are then matched at an arbitrary point $\mu_{_{\rm
M}}$ in the domain $(-1,1)$ following the standard
procedure in the solution of an eigenvalue problem. In
particular, for a trial value of the eigenfrequencies we
look for a zero of the Wronskian evaluated at $\mu_{_{\rm
M}}$, with different zeros representing sequences of
different mode numbers. Once a zero is found, the
eigenfrequency is used in equation~(\ref{eigen_eq}) to
calculate the corresponding eigenfunction. This procedure
is repeated for different degrees of differential
rotation so that a sequence is built describing the
eigenfrequencies of the rotating shell from uniform
rotation up to extreme differential rotation. The results
of these calculations are discussed in detail in the
following Section.

\section{Numerical Results}

	This Section briefly presents the results of the
numerical solution of equation~(\ref{eigen_eq}) for the
two laws of differential rotation discussed above.

\subsection{$j={\rm const.}$ differential rotation}

\begin{figure}[hbt]
\centering\leavevmode
\psfig{file=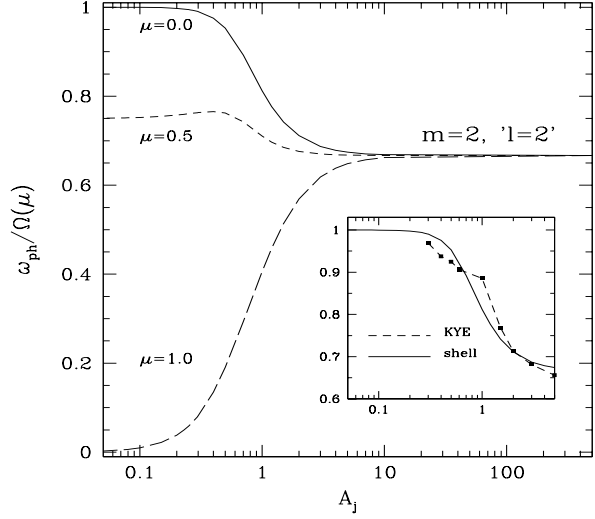,height=7.2truecm,width=8.2truecm,angle=0,clip=}
\caption{Behaviour of the normalized phase velocity
$\omega_{\rm ph}/\Omega(\mu)$ for different values of the
parameter $A_j$ for a $j={\rm const.}$ law of
differential rotation. Different curves refer to the
values of the eigenfrequencies at different latitudes on
the shell. All of the curves refer to a mode which tends
to the $m=l=2$ $r$ mode in the limit of uniform rotation
(because of this we refer to it as an $m=2$, `$l=2$'
mode). The small inset shows a comparison between the
ratio $\omega_{\rm ph}/\Omega(\mu=0)$ obtained with the
present shell approach and the corresponding quantity
obtained by KYE and indicated with filled squares.}
\label{fig1}
\end{figure}

	We show in figure~\ref{fig1} the values of the
phase velocity $\omega_{\rm ph}$ normalized to the value
of the angular velocity at different latitudes on the
shell. (This normalization is particularly convenient as
it allows one to detect immediately whether corotation,
corresponding to $\omega_{\rm ph}/\Omega(\mu) \to$ 1,
takes place or not.)  As expected, in the limit of
uniform rotation, the curves converge to $2/3$, the value
given by the dispersion relation (\ref{dispersion}) for
an $m=l=2$, $r$ mode. For progressively smaller values of
$A_j$, on the other hand, the curves split in response to
the different angular velocities at different angular
positions on the shell. Surprisingly, corotation is
reached only at the equator (i.e. $\mu=0$) and then only
in the limit of $A\to 0$, which is not physically
interesting.

\begin{figure}[htb]
\centering\leavevmode
\psfig{file=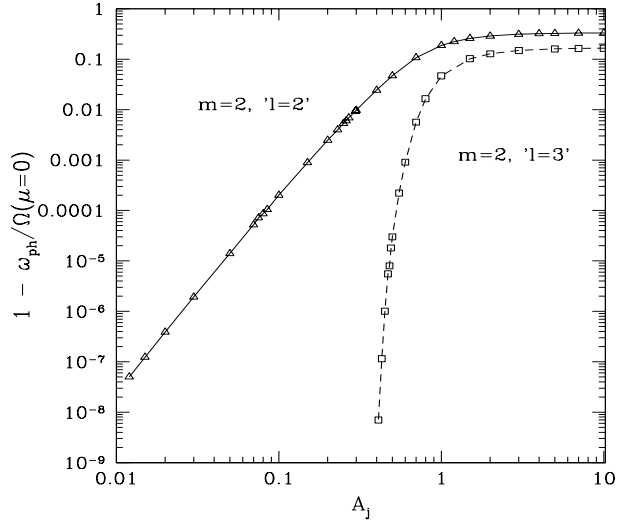,height=7.2truecm,width=8.2truecm,angle=0,clip=}
\caption{Deviation from unity of the ratio of the phase
velocity to the local angular frequency at the equator
(see figure~\ref{fig2}). The small triangles and squares
show the values of the computed eigenfrequencies.}
\label{fig2}
\end{figure}

	This can be clearly seen in figure~\ref{fig2},
where we show the deviation away from unity of the ratio
$\omega_{\rm ph}/\Omega(\mu=0)$.  For both curves, which
correspond in the limit of uniform rotation to an $m=l=2$
and an $m=2$, $l=3$ mode respectively, the behaviour for
small values of the parameter $A$ is well fitted by a
power law $\omega_{\rm ph}/\Omega(\mu=0) \sim K A^n$,
where $n\simeq 3.76$ for $m=2$, '$l=2$', and $n\simeq
53.4$ for $m=2$, '$l=3$' ($K$ is here a positive
constant). As a result, the deviation away from unity
tends to zero only in the limit of $A\to 0$ and we
therefore conclude that {\it no corotation} appears in
these modes in the range of validity of a $j={\rm
const.}$ law of differential rotation.

	The small inset in figure~\ref{fig1} shows a
comparison between the ratio $\omega_{\rm
ph}/\Omega(\mu=0)$ obtained with the present shell
approach and the corresponding quantity (indicated with
filled squares) obtained by KYE with their numerical code
for a slowly rotating stellar model with axis ratio
0.95. There is a rather good agreement, at least for the
range of differential rotation rates in which the mode
calculation was possible. Such an agreement brings
confidence about the relevance of the results obtained
with the shell model also for multidimensional star
models.

	When looking at the results of KYE, it becomes
apparent how the behaviour of the eigenfrequencies might
have suggested the appearance of a corotation point.  As
mentioned in the Introduction, when the degree of
differential rotation increases past a certain threshold,
very large radial and polar gradients appear in the
equations for the eigenvalue problem. The solution of
very stiff equations using finite difference techniques
is a very difficult task and it is therefore not
surprising that KYE were not able to find convergent
solutions to the eigenvalue problem for $A\lesssim 0.3$.
By making use only of an ordinary differential equation,
the shell approach bypasses this difficulty and provides
an accurate solution for any value of $A$.

\subsection{$v={\rm const.}$ differential rotation}

\begin{figure}[htb]
\centering\leavevmode
\psfig{file=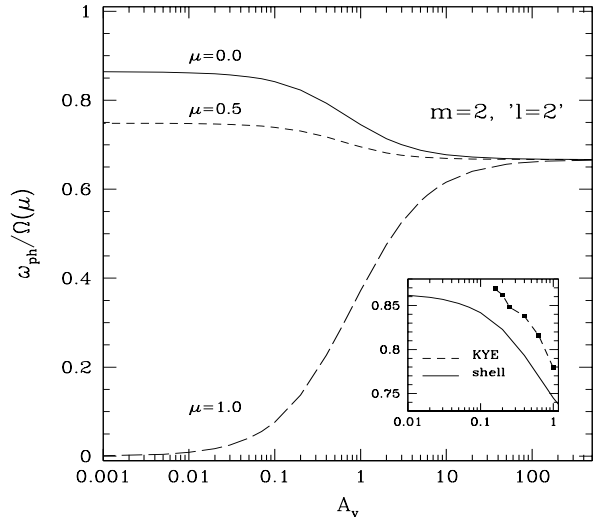,height=7.2truecm,width=8.2truecm,angle=0,clip=}
\caption{Same as in figure~\ref{fig3} but for a $v={\rm
const.}$ law of differential rotation.}
\label{fig3}
\end{figure}

	In the case of the $v={\rm const.}$ law, the
absence of a corotation point is even more evident. In
figure~\ref{fig3} the eigenfrequency is plotted as in
figure~\ref{fig1}. There, we clearly see that the ratio
$\omega_{\rm ph}/\Omega(\mu)$ is below unity everywhere
on the shell. The comparison with the results by KYE,
shown in the small inset of figure~\ref{fig3}, is less
good than the one seen for a $j={\rm const.}$
differential rotation law but the overall behaviour is
rather similar.

	In addition to the differential rotation laws
(\ref{diff_rot_j}) and (\ref{diff_rot_v}) we have also
investigated a quadratic differential rotation law of the
type $\Omega/\Omega_{_{\rm P}} = 1 + (\mu^2-1)/{B}$,
where $B>0$ and uniform rotation is reached for $B \to
\infty$. Also in this case, no evidence for corotation
was found for the values of the parameter $B$ satisfying
the Rayleigh stability criterion (i.e $B>2$).

\section{Conclusion}

	We have used a thin-shell model to investigate
the behaviour of Rossby-Haurwitz waves (the analogues of
$r$ waves on thin shells) in the presence of differential
rotation.  Our simplified approach replaces the set of
partial differential equations for a multidimensional
stellar model with a single ordinary differential
equation that can be solved to arbitrary accuracy. The
numerical solution of the eigenvalue problem has shown
that there is no evidence of corotation, even for
asymptotically large values of differential rotation. As
a result, the eigenvalue problem for Rossby-Haurwitz
waves never becomes singular, as had instead been
suggested by recent calculations of differentially
rotating axi-symmetric Newtonian stars. We have found
that, for this simplified model, $r$ modes can in
principle coexist with arbitrarily large rates of
differential rotation, the eigenfrequencies and
eigenfunctions being suitably modified in response to the
degree of differential rotation.

	The relevance of the results found here for more
realistic stellar models is not easy to assess, although
the equations solved here show many of the mathematical
features of the corresponding equations for Newtonian
differentially rotating stars, or for slowly rotating
relativistic stars.

\ack It is pleasure to acknowledge fruitful discussions
with M. Abramowicz, N. Andersson, S. Bonazzola,
Y. Eriguchi, K. Kokkotas, J. Miller, J. Ruoff and
N. Stergioulas.
Financial support for this research has been provided by
the MURST and by the EU Programme ``Improving the Human
Research Potential and the Socio-Economic Knowledge Base"
(Research Training Network Contract HPRN-CT-2000-00137).

\section*{References}
\begin{harvard}

\item[]Andersson N. 1998 {\it Astrophys. J.} {\bf 502} 708

\item[]Andersson N. and Kokkotas K.D. 2001 to appear in
	{\it Int. J. Mod. Phys.} {\tt gr-qc/0010102}

\item[]Eriguchi Y. and M\"uller E. 1985 {\it
	Astron. Astrophys.} {\bf 146} 260

\item[]Friedman J.L. and Morsink S.M. 1998 {\it
	Astrophys. J.} {\bf 502} 714

\item[]Friedman J.L. and Lockitch K.H. 2001, Proceedings
	of the IX Marcel Grossman Meeting, World Scientific,
	eds. V. Gurzadyan, R.  Jantzen, R. Ruffini; {\tt
	gr-qc/0102114}

\item[]Haurwitz B, 1940 {\it J. Mar. Res.} {\bf 3}, 254.

\item[]Karino K., Yoshida S. and Eriguchi Y. 2001 {\it
	Phys. Rev. D} {\bf 64} 024003

\item[]Kojima K. and Hosonuma 1999 {\it Astrophys. J.} {\bf 520} 788

\item[]Levin Y. and Ushomirsky G. 2001 {\it
	Mon. Not. R. Astr. Soc.} {\bf 322} 515

\item[]Lindblom L., Tohline J.E. and Vallisneri M. \PRL {\bf 86} 1152

\item[]Lockitch K.~H. and Andersson N. 2001, {\it preprint} {\tt gr-qc/0106088}

\item[]Rezzolla L., Lamb F.K. and Shapiro
	S.L. 2000 {\it Astrophys. J.} {\bf 531} L139

\item[]Rezzolla L., Lamb F.K., Markovic D. and Shapiro
	S.L. 2001a {\it Phys. Rev. D} in press

\item[]\dash 2001b {\it Phys. Rev. D} in press

\item[]Ruoff J. and Kokkotas K.D. 2001a, {\it preprint} {\tt gr-qc/0101105}

\item[]Ruoff J. and Kokkotas K.D. 2001b, {\it preprint} {\tt gr-qc/0106073}

\item[]Spruit H.C. 1999 {\it Astron. Astrophys.} {\bf 341} L1

\item[]Stergioulas N. and Font J.A. 2001 \PRL {\bf 86} 1148

\item[]Stewartson K. and Rickard J.A. 1969 {\it
	J. Fluid. Mech.}, {\bf 35} 759

\item[]Yoshida S. 2001 to appear in {\it Astrophys. J.}
	{\tt gr-qc/0101115}

\item[]Yoshida S. and Futamase T., {\it preprint} {\tt gr-qc/0106076}

\end{harvard}

\end{document}